# The neuroconnectionist research programme

Doerig, A.*[1,2], Sommers R.*[3], Seeliger, K.[4], Richards, B.[5,6,7,8,9], Ismael, J.[10], Lindsay, G.[11], Kording, K.[12], Konkle, T.[13], van Gerven, M.A..J.[1], Kriegeskorte, N.[14], & Kietzmann, T.C.[1,2]

[1] Donders Institute for Brain, Cognition & Behaviour, Nijmegen, the Netherlands
[2] Institute of Cognitive Science, University of Osnabrück, Germany
[3] Max Planck Institute for Psycholinguistics, Department of Neurobiology of Language, Nijmegen, the Netherlands
[4] Max Planck Institute for Human Cognitive and Brain Sciences, Leipzig, Germany
[5] Dept. of Neurology and Neurosurgery, McGill University, Montréal, QC, Canada
[6] School of Computer Science, McGill University, Montréal, QC, Canada
[7] Mila, Montréal, QC, Canada
[8] Montréal Neurological Institute, Montréal, QC, Canada
[9] Learning in Machines and Brains Program, CIFAR, Toronto, ON, Canada
[10] Columbia University, New York, USA
[11] New York University, New York, USA
[12] Bioengineering, Neuroscience, University of Pennsylvania, Pennsylvania, PA, USA and CIFAR, LMB
[13] Harvard University, Cambridge, MA, USA
[14] Zuckerman Institute, Columbia University, New York, USA
* equal contributions

Corresponding author: adoerig@uos.de

**Abstract:**

Artificial Neural Networks (ANNs) inspired by biology are beginning to be widely used to model behavioral and neural data, an approach we call *neuroconnectionism*. ANNs have been lauded as the current best models of information processing in the brain, but also criticized for failing to account for basic cognitive functions. We propose that arguing about the successes and failures of a restricted set of current ANNs is the wrong approach to assess the promise of neuroconnectionism. Instead, we take inspiration from the philosophy of science, and in particular from Lakatos, who showed that the core of scientific research programmes is often not directly falsifiable, but should be assessed by its capacity to generate novel insights. Following this view, we present neuroconnectionism as a cohesive large-scale research programme centered around ANNs as a computational language for expressing falsifiable theories about brain computation. We describe the core of the programme, the underlying computational framework and its tools for testing specific neuroscientific hypotheses. Taking a longitudinal view, we review past and present neuroconnectionist projects and their responses to challenges, and argue that the research programme is highly progressive, generating new and otherwise unreachable insights into the workings of the brain.

**Introduction:**

While the study of cognition is a millennia-old endeavor (e.g., already present in Aristotle's *De Anima*), recent years have seen remarkable advances in both experimental and computational analysis techniques, yielding more powerful ways in which we can study and model computations in the brain[1]. Yet, the level at which cognition should best be understood remains a hotly debated topic. Modeling biology at the molecular level may not guarantee a deeper understanding of the core principles of cognition, any more than one brain can serve as an explanation of another. Instead, the task of cognitive computational neuroscience is to find the right level of abstraction, close enough to biology to preserve the essential mechanisms, but abstract enough to discard details not required for cognitive function, while reproducing the trajectory from actively sensed input, through internal representations realized in neural processes, to complex goal-directed behaviors[2].

Traditional experimental approaches, for example adopted in visual neuroscience, often operate at the rather coarse-grained explanatory level of contrasting experimental conditions. For instance, by running experiments with highly controlled stimuli, neural firing rates have been interpreted in terms of category selectivity: neurons are deemed selective for "faces", "houses", or "tools"[3–6]. This approach has merit. Its controlled settings allow for maximal interpretability and suggest a clear taxonomy of neural selectivity[7]. Yet, human-interpretable labels for neural activity are limited by the researchers' imagination, or simply by labels that language makes available to us. Nature does not owe us easy answers: neural selectivity may often rely on more complex features that only imperfectly map onto human-interpretable categories[8]. In addition, showing selectivity for a higher-level concept does not provide an answer as to how the brain may have arrived at this high representational level from noisy, lower-level sensory input data.

Together, these observations highlight the need for neuro-computational models that can bridge explanatory and computational levels and predict neural data and behavior while being grounded in sensory data. This enables answering central questions of cognitive neuroscience: How can we link sensory input to neural data across brain regions, not only at the level of individual cells but also at the population level? How can neural processes be linked to behavior? How do neural representations change, not only through space, but also through time (from milliseconds to developmental trajectories)? How can past experience be encoded in the system and which types of feature selectivity allow for task-general robust performance?

In short, for a complete picture of how the mind works, neuroscience is in need of computational models that go beyond the limits of human-interpretable labels and tie together multiple levels of explanation. A good model of how some cognitive function is implemented in the brain should therefore (non-exhaustive list):

   a. Specify which computations are carried out by the brain (computational level).
   b. Show how these computations lead to complex behavioral patterns that can be tested in experiments (behavioral level).
   c. Show how these computations lead to complex neural dynamics that can be tested in experiments (single unit and collective dynamics levels).
   d. Show how these computations can be carried out in complex settings, beyond simplified experiments (rich domain knowledge).
   e. Show how these computations can be grounded in natural sensory information, rather than high level features provided by humans (sensory grounding).
   f. Show how these computations arise from adaptive processes that unfold at multiple timescales (inference, learning, development and natural selection).

To address these desiderata in a consistent explanatory framework, researchers have turned to computational modeling using Deep Neural Networks[9,10] (DNNs; the latest incarnation of Artificial Neural Networks[11] (ANNs)). ANNs seem well-suited to tackle the challenges described above[12]. First, ANNs are made of simple units that collectively implement complex computations that drive the network's behavior. That is, they offer a framework spanning the single unit, collective dynamics, behavioral and computational levels (desiderata a-c). Second, ANNs use millions (sometimes billions) of synaptic parameters to encode rich domain knowledge, while optimizing connectivity over time to learn (desideratum d). Third, ANNs are grounded in sensory input, meaning that they can be trained on raw "sensory" data to fulfill "behavioral" needs, without the need for human-engineered input features, offering a link between sensation, cognition and action (desideratum e). Finally, by allowing us to iteratively compare different biologically plausible learning rules and objectives, ANNs can help uncover how learning and cognitive development are made possible, and how they interact with architectural network features (desideratum f). Importantly, the architectural flexibility of ANNs and the different ways in which they can be trained allow for explorations, in a top-down fashion, of which biological details are needed to satisfy desiderata (a)-(f) for a given cognitive phenomenon. That is, they allow for complex hypothesis testing, much like traditional experimentation.

These characteristics allow researchers to rigorously test ANNs against large-scale behavioral and neural experimental datasets collected from a large array of brain regions[13], and adjust the level of biological detail where needed - an approach markedly different from classic machine learning (ML) engineering, which is usually geared towards high performance on a small number of benchmarks. The approach furthermore goes beyond classic connectionist models of the 20th century[14], which used smaller networks to explain higher-level cognitive tasks rather than the more recent focus on multi-level understanding of brain function. Due to the close integration into neuroscience, both in terms of network design and mapping of internal representations to brain function and neural data, we term this approach *neuroconnectionism*.

Neuroconnectionism has already been fruitful in a wide variety of neuroscientific settings, including vision[15–21], audition[22,23], semantics[24–26], language[27,28], reading[29], decision making[30–33], memory[34], game playing[35], motor control[36–39], and the formation and coding principles of brain areas[40–45]. See section *The neuroconnectionist belt* for more details, and [46–52] for reviews[1].

At the same time, neuroconnectionism does not remain unchallenged. It has been raised that ANNs differ strongly from biology, that they often behave in non-human ways[53–56], and that the complexity of the models prohibits true understanding[54,55]. These challenges are sometimes interpreted as suggesting that ANNs are not useful models for learning about the brain. For example, an argument might proceed like this: ANNs are sensitive to adversarial attacks, unlike humans; therefore, these models have been falsified and thus should not be used as models of the visual system. However, this logic only follows from a Popperian view of science (Box I).

Instead, we argue that these challenges are better seen as signposts. This changes the interpretation of the above example: ANNs are sensitive to adversarial attacks, unlike humans; therefore, they must lack the relevant mechanisms that yield a more robust, human-like visual representational format, which will be important to improve future models. This conclusion follows a more Lakatosian philosophy (Box I). To elaborate on this view, we introduce neuroconnectionism as a Lakatosian research programme, which, with its explicitly longitudinal perspective on the developments in the field, offers fruitful grounds for discussion. This provides a new view on the challenges mentioned above, which act as signposts towards further developments.

---

[1] Please note that the authors of this piece predominantly work in the sensory domain, in particular in vision. The provided examples are hence biased towards this research field. The larger neuroconnectionism community is however not limited to sensory processing, but involves, among many other areas, language processing, memory, (meta) learning, as well as movement and embodiment/robotics.

**Neuroconnectionism as a Lakatosian research programme**

In philosophy of science, Lakatos[57] proposed a general framework to evaluate scientific approaches. According to his view, science is typically carried out within research programmes. Such programmes share a *hard core* of background assumptions that are not typically challenged from within the programme, and contain a *belt of* auxiliary hypotheses that are experimentally tested. While the core cannot be altered without abandoning the research programme, elements of the belt are and should be subject to change.

Given these two elements, core and belt, the value of a research programme is determined not just by its current experimental success relative to other research programmes, but also based on whether it is *progressive* rather than *degenerating* - an explicitly longitudinal perspective. Progressive research programmes generate new theoretical insights and novel predictions where some of these predictions are corroborated by empirical findings. Degenerating research programmes do not have these two characteristics: they often lack new theoretical developments and devolve into repeated corroboration of very similar ideas.

In the following, we will characterize and evaluate neuroconnectionism as a Lakatosian research programme composed of a set of auxiliary hypotheses in a belt centered around a unifying core, rather than a single falsifiable hypothesis. That is, we consider productive ways to use ANNs to gain insight into understanding the mind and brain (rather than considering any specific current model as *the* model of the brain that is true or false). We will describe the tools that are currently used for hypothesis testing in the field, review past challenges and successes, and revisit current controversies to evaluate whether neuroconnectionism is progressive.

Importantly, our arguments hold even if one does not accept the Lakatosian view on progress in science. While the Lakatosian distinction between core and belt is particularly well-suited to contribute to the ongoing debate about the merits and weaknesses of neuroconnectionism, what matters is that scientific theories always have core guiding principles that are relatively isolated from direct empirical testing - an uncontroversial view in the philosophy of science (see Box I). These core principles define directions of inquiry, including which experiments are conducted, and what is taken to corroborate or weaken theories. Theories are hence judged through a holistic assessment of their successes and fertility in guiding experimental pursuits. The Lakatosian perspective that we adopt is merely one helpful way of expressing these general ideas.

**BOX I: THEORY SELECTION & PHILOSOPHY OF SCIENCE**

Through much of the twentieth century, the Popperian view that theories are rejected when they are falsified in tests was dominant. A scientific theory generates predictions and tests are run to see if the predictions are correct. If they are not, the theory is falsified and can be rejected (it may take more than one test to show that the failure wasn't experimental error, but once the result is accepted, the theory must be rejected).

The Popperian view assumes that the logic of disconfirmation follows the following schema, where T is a theory, and O is an observation.

*If T, then O*

*Not O*

*Hence, not T*

It was noticed first by Pierre Duhem (the French theoretical physicist and historian of science) in 1906, and later reinforced by Quine (one of the most influential analytic philosophers in the twentieth century), that science does not work like that in practice and could not work like that in principle. One typically (and perhaps always) needs to combine the hypothesis to test with auxiliary beliefs to extract empirical predictions. When it is made fully explicit, the logic looks as follows:

*If T, and A1, and A2, and A3,…, and An, then O*

*Not O*

*Hence, not T, or not A1, or not A2, or not A3,…, or not An*

Where (A1, and A2, and A3,…, and A$n$) are auxiliary hypotheses needed to generate predictions.

For this reason, it is never a single hypothesis, but a whole collection that generates predictions, any one of which might be at fault if the prediction is not vindicated.

This fact about the logic of confirmation has been one of the centerpieces of twentieth century philosophy of science. Quine popularized the idea that a theory forms a web of beliefs related by inferential connections. The web has a topology with beliefs that can be most directly subjected to empirical tests at the periphery, and others that are insulated from direct testing by long chains of intermediary hypotheses. Beliefs at the periphery describe localized observable facts; beliefs at the center describe the kinds of general beliefs that guide the explanation of a whole body of phenomena. These are highly connected in the web and separated from empirical predictions by mediating propositions. One could hold onto the general beliefs in the face of mounting evidence if one was willing to make adjustments in other parts of the web.

Lakatos' distinction between core and belt acknowledges this holistic nature of confirmation, while offering a pragmatics of testing that is well-suited to capture how science actually works. In the Popperian view, testing is a matter of checking whether a theory accords with fact. On a Lakatosian view, it goes hand in hand with the development of theory. One adopts a set of core

> principles as a kind of working hypothesis, proceeding on the assumption that they are correct and using them to try to understand the phenomena. Testing is a process that involves striving to bring theory and fact into closer agreement by exploring ways in which the core principles can be preserved while accommodating the evidence. If the core principles are held fixed and leeway is granted to explore alternative auxiliary hypotheses, testing can be directed at the belt, giving the theory every chance of preserving the core while accommodating the evidence. This is a way of targeting disconfirmatory evidence against one part of the web. A theory is rejected not as the result of a direct conflict with the evidence, but because the attempt to preserve the core principles becomes so cumbersome that they cease to form a productive working hypothesis.
>
> For example, in astronomy, there were deviations in planetary trajectories from the smooth ellipses predicted by Newtonian mechanics. Instead of rejecting Newtonian laws, people interpreted this as a reason to look for ways to explain the distortion, assuming the correctness of the laws (e.g., an unseen planet). Hence, a belt claim was falsified but the core was not abandoned. Instead, the core is changed only when it becomes unproductive to hold onto it. For example, in the twentieth century, evidence accumulated against Newtonian celestial mechanics that could not be solved assuming the correctness of the laws, which led to its rejection and the development of general relativity.

*The neuroconnectionist core*

As discussed in the introduction, ANNs hold promise to fulfill desiderata (a)-(f) by tying together multiple levels of explanation, incorporating rich domain knowledge, and going beyond the limits of human-interpretable labels. In a sense, neuroconnectionism can be seen as an attempt to address these desiderata by using ANNs, which motivates why ANNs are at the core of the research programme. In this section, we elaborate on this further and describe the core tenets of neuroconnectionism and how they relate to specific properties of ANNs.

ANNs consist of interconnected sets of rather simple units, which perform local computations and which by themselves have no perceptual or cognitive abilities. Instead, perceptual or cognitive computations emerge as a collective property of the network, distributed across many units. Information flow through the network determines its function, and is itself determined by the connection strength between the many units. How to set up and weigh these connections to encode domain knowledge and perform according to a set objective is a difficult problem: for a fully connected recurrent ANN with N units, there are $2^{N \times N}$ possible structural connectivity patterns, under the simplifying assumption of connections being either 0 or 1[2].

Because of this complexity, researchers constrain the connectivity motifs in the form of network *architectures* that determine which connections are possible - providing strong inductive biases. Most current network architectures organize their units in layers and within-layer-connectivity is determined as part of the architecture. In feedforward networks, units of a given layer only receive inputs from the layer below and only send outputs to the layers further above. Recurrent networks allow for additional top-down and lateral message passing. Another noteworthy architectural aspect to constrain the number of learnable parameters in a network is to make it convolutional[16,58,59]. This convolutional architecture was directly inspired by seminal findings in visual neuroscience[60]. Convolutional layers are subdivided into feature maps. Each unit in a feature map has its own receptive field, but all units share the same selectivity. Applied in the visual domain, this means that when the network learns to detect a feature at one location, it automatically generalizes to other locations. This is a useful inductive bias for vision: since the same object can occur anywhere in an image, it is sensible to look for the same features everywhere, without having to re-learn them at each location. Other domains, such as olfaction, do not share this property and thus favor other inductive biases. Many different architecture types exist in the literature, but the above examples illustrate the main point: by constraining connectivity, each architecture not only reduces the number of parameters in the network, but also implements inductive biases that determines which functions can be efficiently learnt.

---

[2] For a system of 8 units, that is 18 quintillion 446 quadrillion 744 trillion 73 billion 709 million 551 thousand 616 connectivity patterns.

While architectural design defines the general space of possible solutions, the individual network parameters still need to be optimized. For example, ResNet-50, a popular convolutional architecture, has over 23 million parameters[61]. To learn these parameters, networks are typically presented with input *datasets* (e.g. natural images, auditory signals, or text corpora) and their target behavior is defined based on one or several mathematical *objectives* (also called loss functions). Ultimately, several low-level objectives may be subsumed in higher level objectives, such as "fitness". Deviations from the target, or errors, are then minimized by updating the weights between network units, which is determined by a *learning rule*. To do so, the weights of each individual unit must be changed in accordance with the contribution of the unit to the whole network's error. Given that ANNs can have millions of units (operating over extended time in the case of recurrent networks), finding good learning rules is far from trivial. How to attribute the contribution of each unit to the overall error is called the *credit assignment* problem. The most widely used learning rule to solve credit assignment in ANNs is backpropagation[62], but there are many others too, e.g. [63,64]. It is worth noting that each learning rule comes with its own hyperparameters, such as a learning rate and how the learning rate changes over training. Overall, many datasets, objectives and learning rules are used in neuroconnectionism, and they strongly impact the behavior and internal states of ANNs (see section *The neuroconnectionist belt* for details).

The approaches described above for designing and training ANNs offer ways of testing hypotheses of how certain functions may be implemented in the brain while enabling researchers to perform large-scale in-silico experiments[46,49–51]. Indeed, an ANN's architecture, dataset statistics, objectives, and learning rules can be mapped onto central questions of cognitive neuroscience, allowing computational traction to untangle the interacting contributions of pre-specified structure (for ANNs: determined by the architecture) and experienced input (for ANNs: training dataset), why neural selectivity in a given brain region is the way it is (for ANNs: which objectives are being optimized), and how the brain may adjust its internal representations (for ANNs: credit assignment / learning rules). All of these questions can be studied across levels of explanation and temporal scales, incorporating rich domain knowledge grounded in sensory data, in line with desiderata (a)-(f).

Importantly, there is likely no shortcut to the iterative learning and processing occurring in neural networks. Many complex iterative processes – such as cellular automata like Conway's Game of Life or Rule 110 – can be specified on half a post-it note, yet their eventual behavior after many steps is fundamentally unpredictable[65,66]. Similarly, the learning and processing dynamics of neural networks are unpredictable despite having simple mechanics. This unpredictability has been proven for recurrent neural networks[67]. Put differently, ANNs need to be processed iteratively to study the effects of training, architecture, task, and dataset. No analytical mathematical model exists to date that could use shortcuts to directly get the final result without going through the iterative process. The same is likely true of the brain, a complex, deep, massively parallel and recurrent neural network. Therefore, it likely can only be modeled using another complex iterative process, such as ANNs, while performing careful hypothesis testing at a level of abstraction that offers the best insights to the process at hand (the optimal level of abstraction may depend on the phenomenon under investigation). This lack of simpler models justifies the use of complex ANN models in neuroconnectionism.

With this information in place, we can state a Lakatosian core of the neuroconnectionism research programme (see also Figure 1):

> ***The core of neuroconnectionism is that brain computations, representations, learning mechanisms, and inductive biases are best understood via modeling in ANNs that are defined by their architectures, input statistics, objective functions and learning rules. Multilevel and sensory-grounded iterative models like ANNs are needed to model multilevel and sensory-grounded iterative brain computations, and thereby achieve desiderata (a)-(f).***

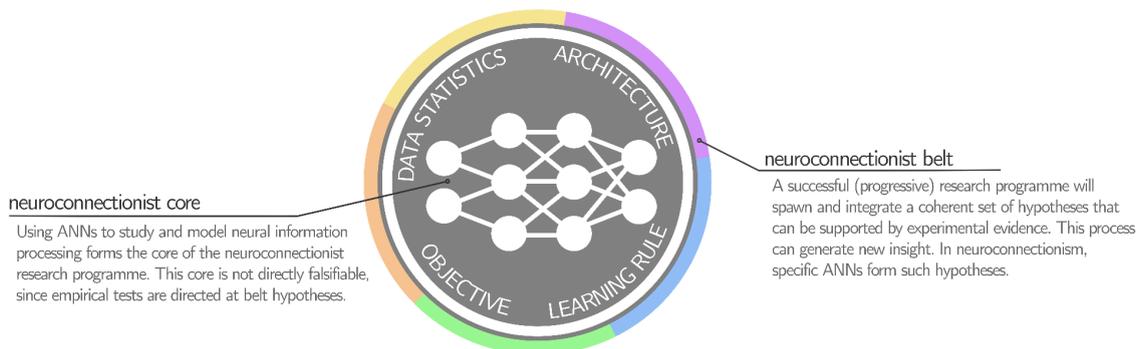

**Figure 1** | Illustration of the neuroconnectionist research programme. The core of neuroconnectionism (depicted as the gray center) is to use ANNs, defined by their architectures, objective functions, input data statistics and learning rules, to test neuroscientific hypotheses. Following the Lakatosian view, experimental investigations are not directed at the core of the research programme, but at a surrounding belt of auxiliary hypotheses (depicted as coloured arcs, e.g., the green arc could stand for the hypothesis of CNNs as models of the human vision). Falsifying individual belt hypotheses does not invalidate the research programme as a whole, but instead calls for improving the belt. Hence research programmes are not judged by attempting to falsify the core directly, but rather by assessing if they are progressive or degenerating. Good research programmes are progressive: they successfully integrate hypotheses into a coherent belt, produce new hypotheses, and make testable predictions that are corroborated.

The above definition aims to capture the core ideas shared by the many projects that contribute to neuroconnectionism. However, like all research approaches, there can be no rigid definition. Still, just like how members of a family do not share *all* traits, but still possess a family resemblance[68], so too do members of the neuroconnectionism research programme constitute a diverse but cohesive family of approaches. Distinctive features of neuroconnectionist models include:

*Explaining cognition as primary explanatory goal*: Neuroconnectionist models are primarily aimed at explaining cognitive functions, not at describing biology with the highest possible detail. Biological detail is added to models in a top-down, hypothesis driven fashion, when these details are necessary for explaining behavioral or neural data. This makes neuroconnectionism different from approaches aiming to perfectly replicate a human brain in silico and from biophysical models aiming to model every aspect of a neuron.

*Distributed representations and computations*: The modeled property emerges from the collective behavior and dynamics of simple units which, taken independently, do not show the modeled property. This distributed nature is central to neuroconnectionism, since it naturally bridges between explanatory levels, from single units, through collective dynamics and onto behavior. This distinguishes neuroconnectionism from traditional models where each parameter has a straightforward interpretation, such as drift-diffusion models, and requires special interpretation frameworks to cope with the distributed nature of computations (see the *neuroconnectionist toolbox* section).

*Iterative training & inference*: The high-dimensionality of distributed models makes it virtually impossible to tune all parameters by hand, so iterative training is required. At inference too, the behavior and dynamics cannot be simplified into a simple interpretable equation: to know what the model predicts, one needs to run the model. This iterative nature distinguishes neuroconnectionism from approaches that can be formulated in a small number of non-iterative equations, such as (hierarchical) Bayesian models.

There are of course many edge cases. For example, while grounding in sensory input is a desideratum of the neuroconnectist approach, we note that not all neuroconnectionist models are grounded in sensory input. In language or memory models, for instance, the input to the model often consists of high-level concepts, such as words. Still, we consider them neuroconnectionist, since the sensory nature of the inputs may not be relevant in these particular cases, and could be added to later models if needed. As another example, certain neuroconnectionist models may not explicitly employ a "behavioral" objective function, but rather attempt to minimize energy efficiency (e.g. [69]). End-to-end learning where models are directly fitted to neural data instead of trained on a behavioral task are also neuroconnectionist since they are a way to test hypotheses about which architectures are capable of reproducing neural dynamics[19]. In addition, models do not need to model cognition in general, but can focus on specific components. Understanding which aspects of cognition can only be modeled jointly is part of the hypothesis testing process. For example, sensory-motor interactions in embodied models may or may not be required to explain certain aspects of visual processing. Lastly, models that are architecturally different from the brain, such as GANs or Transformers, can also be used to test neuroscientific hypotheses and thereby contribute to neuroconnectionism[70–72]. In short, most ANN models that can be used for neuroscientific hypothesis testing can be considered neuroconnectionist given a researcher's motivation of how the model could be rooted in biology, and how it tests theories about neural computations.

Importantly, neuroconnectionist approaches are about delineating mechanisms and theoretical understanding through hypothesis-driven research, which differs from the goals of engineering. Around the time of the first large-scale vision networks[58,73], neuroconnectionist models were borrowed directly from engineering applications. For example, it was found that the engineering models that performed the best on engineering benchmarks (as of 2014) were also better at predicting brain activity[21]. This may have led to a form of "computational opportunism" in which researchers could directly test ML models against neuroscientific data without having well-formed hypotheses. However, this trend has come to a halt with more recent network architectures where better task performance does not align with improved predictions of neural firing rates[13]. Indeed, when taking the hierarchical nature of the visual system into account, recent years have seen a constant *decrease* in agreement between ANNs and neuroscience[74]. These trends highlight that focusing on engineering goals based on performance alone is not sufficient for a neuroscientific understanding. Neuroconnectionism needs to develop its own models and metrics.

Following Lakatos' definition of research programmes, the different neuroconnectionist projects all use ANN models as core to their approach. Experiments are not directly applied to the core, but rather to the surrounding belt hypotheses, which are realized via ANN instantiations and tested using various techniques. The promise of the overall research programme is then to be judged based on its ability to generate new insights, corroborate belt hypotheses, derive new testable predictions, and, perhaps most importantly, productively address criticisms to make progress.

*The neuroconnectionist toolbox*

To implement and evaluate the neuroconnectionism research programme, tools are needed to instantiate and test models in its belt. These include tools to test to which extent neural network model instances align with neural data and behavior (Figure 2). Empirical findings revealing areas where ANNs fall short are an essential part of the framework and approach, as they provide crucial data points indicating where our understanding and models need to be improved.

*Network design and training*

A central aspect of the research programme is the creation and training of ANNs. For this aspect, engineering has been instrumental, as it provided researchers with key hardware and software technologies that enable training of task-performing models.

To design ANNs, researchers can choose from a large zoo of network architectures and unit types. The different unit types are abstracted away from the brain to various extents. For example, common computational units go from very simple rectified linear summation (ReLU), to complex units modeling basic memory (e.g. LSTMs). Importantly, researchers have started integrating various aspects of biology into network design while testing them against neural and behavioral data[13,75]. This includes recurrent connectivity[19,76–82], and different base-units, such as richer rate-based neurons[83–85], spiking neurons[86–89], and neurons with multiple compartments[80,90,91].

To train ANNs, researchers need to choose datasets, objectives and learning rules. Many large-scale datasets are openly available. Typically, external datasets are used (i.e., the dataset is the input to the network and the network's output is trained to match the labels). For example, inputs could be images and outputs could be captions giving a semantic description of the image. In addition, networks can also be directly trained to mirror brain activity[19,77,92–99], leveraging recent efforts to record large neural datasets[100–105]. The dataset determines the input statistics that the network learns from, so an important avenue of research is to develop more naturalistic datasets (e.g. [106]).

A large variety of objectives is used, including supervised (e.g. classification, scene captioning), unsupervised (e.g. contrastive learning[107,108], predictive coding[109,110], image generation[44,111,112], temporal stability[113–117], and energy efficiency[69]) and behavioral reward[30,35,118]. The objective determines what the networks try to learn based on the input statistics. Different objectives may for example lead to modeling different brain areas[119]. As another example, one may test whether the ventral visual stream is better modeled using an object categorization or a semantic objective[24], or test the impact of noise in the dataset[120].

The most common learning rule is, by far, backpropagation[62]. In its standard form, backpropagation is not biologically plausible, and a lot of work seeks more plausible versions[91,121–124]. Other learning rules exist too[63,64,125]. Each algorithm has its own characteristics. For example, gradient descent tends to learn the input features with most variance first[126] and produces efficient neural codes[127]. Hebbian learning, on the other hand, is a simpler local rule that has been directly observed during learning in biological systems[128].

In summary, by using frameworks optimized for big data and high throughput, computational neuroscientists can create ANNs and test which biological aspects are necessary to reproduce brain function and behavior. ANNs thereby offer a top-down approach to understanding which aspects of the brain are central to its function.

*Model testing*

Although critics sometimes claim that we cannot understand ANNs[54,129], ANNs are not black boxes. Rather, they are transparent boxes with easy access to each unit's activities and connections. Researchers can perform in-silico experiments at speeds that are orders of magnitude faster than on any biological brain, and, for now, free of ethical concerns that come with classical experimentation. Several tools have been developed to improve the interpretability of ANNs and test them against biological data.

Hypothesis testing in neuroconnectionism relies on testing trained ANNs against brain data on various levels, from neural data up to behavioral patterns. Because of multiple realisability, systems that are structurally very different from the brain can nevertheless enable successful predictions of human behavior, or of the activity in certain brain areas (e.g. [130,131]). Hence, to ensure that an ANN implements a given cognitive function in a human-like way, it needs to map onto human brain processing across levels, ranging from behavioral to neural data, ideally while taking into account the physical constraints that the brain faces, such as, for example, metabolic costs[69,132–134] and considerations of wiring length[45]. Importantly, no single method is perfect, and various complementary approaches are needed. Hence, developing good metrics to compare humans and ANNs across levels is an essential part of neuroconnectionism.

*Behavioral agreement:* At the behavioral level, the outputs of ANNs can be compared to human responses in several settings. Coarse measures such as overall task performance are useful but only of limited precision, since ANNs are becoming increasingly good at tasks for which humans were until recently deemed the gold standard. These include object recognition[58,61,135], board games[136] or video games[137,138]. To overcome these limitations, several more fine-grained methods exist. These include the use of diagnostic readouts to characterize the information represented in a population of units and to translate it into behaviorally relevant measures such as reaction times[139], detailed analysis of error patterns[140], testing on out of distribution examples[141–143], and reproducing psychophysical results that target specific aspects of processing[76,144–150]. See [151–154] for further discussion about how to compare human and ANN behavior. A unified model addressing years of psychophysical experimentation is an important target, yet to be achieved.

*Agreement with neural data:* At the neural level, the activity patterns of ANNs can be compared to biology in several ways. One common way of doing so is using Representational Similarity Analysis[155,156] (RSA). RSA characterizes the internal representations of a system by quantifying the dissimilarities between the population responses to various stimuli. Internal representations of ANNs and the brain are deemed similar if the corresponding response geometries agree. This approach has the benefit of side-stepping the need to directly map from individual ANN units to individual neurons or voxels. It is therefore well suited for comparing ANN activations with neural data, e.g. obtained via neuroimaging or electrophysiology. RSA can be applied using all units of a whole network, or all units of a network layer, units in a feature map, or individual ANN units. The representational geometries measured by RSA can be compared directly between different brain regions and/or ANN layers[106], or they can be combined using linear reweighting to optimally map onto each other[18,157,158]. Interestingly, network units can be multiplied with a constant factor while not affecting network output if the synaptic connections to downstream units are downscaled accordingly. This implies that networks with identical behavior populate a whole subspace of solutions, which still differ in their alignment with brain data if no reweighting is applied. An important challenge for current RSA methods is that they sometimes fail to discriminate between different network architectures, which all perform similarly on this metric[158]. Recent work has started addressing this issue by improving current RSA methods[159] and clarifying which aspects of brain computation should be targeted by RSA[160].

In addition to RSA, which is predominantly aimed at characterizing responses at the population level, encoding models can be used to predict the activity of single neurons or voxels across a range of conditions[21,161,162]. Here, the activity of each biological unit (neuron or voxel) is predicted as a linear combination of ANN unit activations. To prevent overfitting, the underlying GLMs are typically regularized. More recent encoding methods separate spatial from channel dimension activity[96,98]. In their original form, encoding models are not constrained as to which units can explain which biological counterpart. In principle this implies that, for example, higher-level brain regions can be explained by lower-level network features, or that thousands of cells in the brain can be explained by the response of a single network unit with broadly similar selectivity. New developments are underway to include an ordered hierarchical mapping from ANNs to the brain[74,163], providing a more rigorous test of the alignment between brain representations and ANN models.

A third approach, which builds on encoding models, is to use ANNs to design stimuli that strongly activate a single target biological neuron or area[164–166], allowing for a causal interaction between ANNs and biological brains. This approach therefore goes beyond the otherwise correlational RSA and encoding model approaches.

*In-silico electrophysiology:* In addition to estimating the level of agreement between ANNs and biological brains in terms of behavior and neural recordings, ANNs themselves can be experimented on to better understand their inner workings. Since we have immediate access to all units, their activities and connectivity, almost any experiment is possible. This includes reliance on network initialisation[167], representational similarities across network architectures[159], and tests for the emergence of brain-like computations[69], selectivity profiles[41,45] and cell types[168]. To this end, standard neuroscientific methods can be applied, such as searchlight decoding, measures from signal detection theory, tuning curve analysis, and many more. In addition, ANNs can be selectively lesioned to test the impact of different parts of the network on its ability to map onto brain function. For example, the effect of recurrent connections can directly be assessed by ablating them[19,82,147]. In-silico lesion studies are of course not limited to analyses of networks on their own, but can also be used to evaluate the changes in agreement between network and neural or behavioral data. Lastly, the ability to replicate topographic elements of brain organization can be tested in ANNs to better understand the origins and functional implications of such representational arrangements. First forays in this direction are well underway[41–43,45,169,169,170].

*Developmental agreement:* All of the above methods can be applied at different points during network training, from untrained to fully trained models, and these learning trajectories can be compared with different stages in biological development[171]. Although it is currently unclear which aspects of learning in ANNs are better seen as corresponding to learning during evolution and which are better seen as modeling learning during an organism's lifetime[172], both can be addressed experimentally. For example, the order in which children learn different words can be predicted by the performance of ANNs trained on visual classification and captioning tasks, over and above the expected effects of word frequency[173].

In summary, neuroconnectionism has at its disposal a large array of techniques to train and evaluate neural networks as models of brain computation. Evaluating and contrasting different models for their ability to explain brain data are vital aspects of the research programme as they enable rigorous hypothesis testing, which brings developments in the belt of hypotheses, as discussed next. The fact that each model can be extensively tested from single neurons to behavior is a strong asset of neuroconnectionism that arguably no other approach can claim.

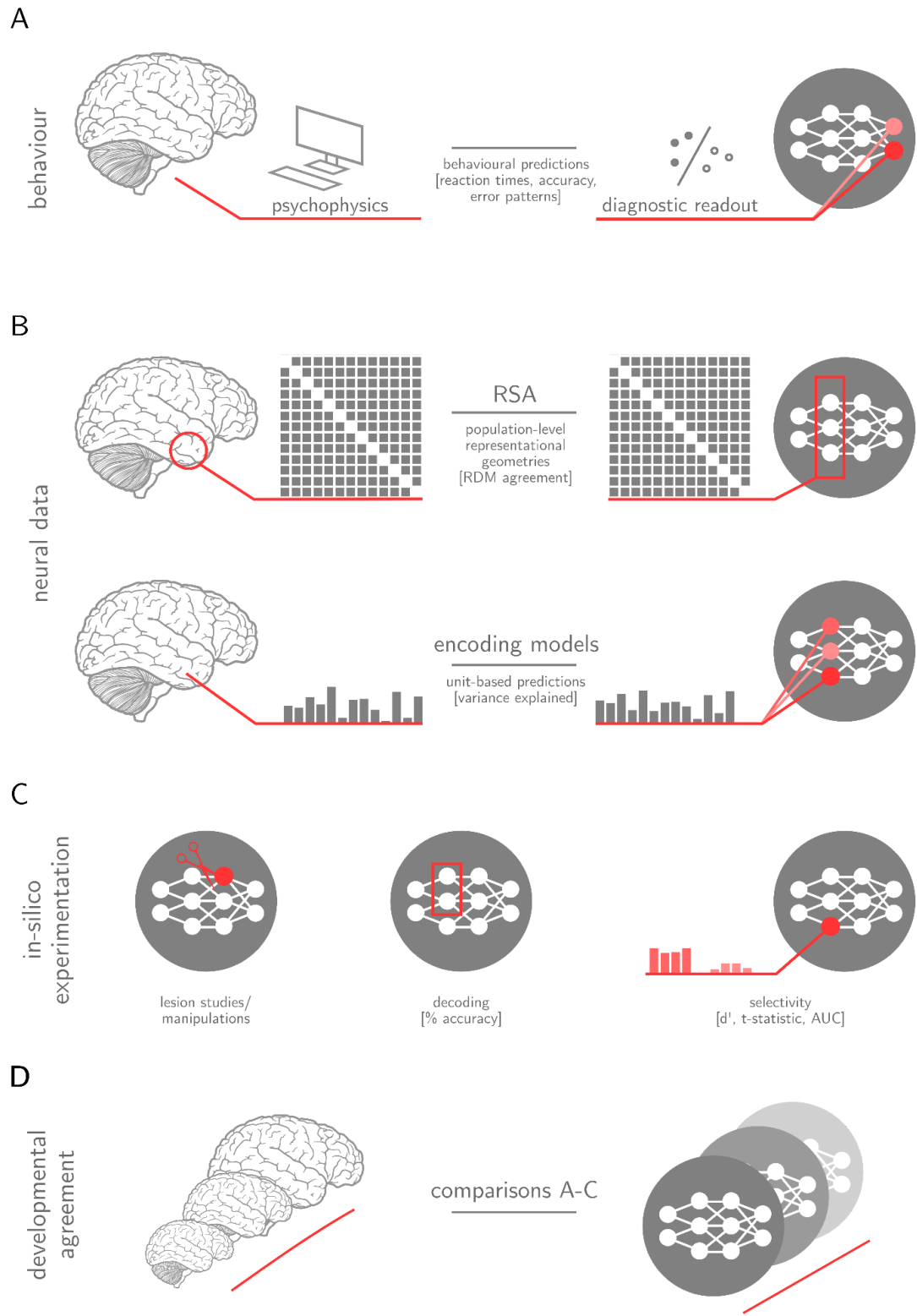

**Figure 2** | The current neuroconnectionist toolkit for model testing. Similar to the belt of a research programme, its toolkit is under constant development. Next to improvements of existing techniques (which historically have focused heavily on sensory processing), this involves new approaches to widen the scope of phenomena that can be investigated. These methods can be applied to compare ANNs with any species (human, primate, mouse, …). **A** Behavioral agreement. Outputs of ANNs are compared to human responses in

diverse settings, such as classification errors and accuracy, reaction times, action patterns, and others. **B** Agreement with neural data. Presenting identical stimuli (input) to the brain and computational model, the recorded brain activity patterns are directly compared to ANN layer activity patterns. The most common methods are RSA (comparing representational geometries) and encoding models (pointwise linear regression). **C** In-silico electrophysiology. ANNs are studied as in-silico models of cognitive functions with standard neuroscientific methods, such as manipulations and lesions, information decoding, unit-based tuning functions and others. Effects of design choices such as recurrent connections, or manipulations such as detailed lesioning patterns can be studied extensively in this setup, something not possible in-vivo. **D** Developmental agreement. Comparing the above at different stages of training to stages of learning in biological brains permits insights into cognitive development. Examples include behavioral patterns, map formation, or changes in neural selectivity with visual experience.

### *The neuroconnectionist belt*

We have now defined the core of the research programme, as well as tools that are used to design, train, and evaluate ANN models across various levels of explanation. This brings us to the belt of the research programme, a set of hypotheses which are to be tested and which evolve as new evidence is integrated. Individual elements of the belt are important, but a more central aim, when taking a Lakatosian perspective, is an evaluation of longitudinal developments (theoretical and empirical), which determine the state of a research programme as either progressive or degenerative. By definition, the belt of a research programme is subject to change. New hypotheses can be derived, and existing hypotheses can be corroborated, altered, and rejected. An individual belt hypothesis that is rejected does not refute the core assumptions upon which a programme is built, but rather provides an important datapoint for future developments. According to this view, the overarching question becomes: How does the neuroconnectionism research programme fare in terms of productivity? Does it generate new insights, and does it address existing challenges? Are there challenges that cannot be overcome by the research programme (roadblocks: rendering it degenerating), or are challenges rather signposts towards open questions that can be addressed by improving the research programme (progressive)?

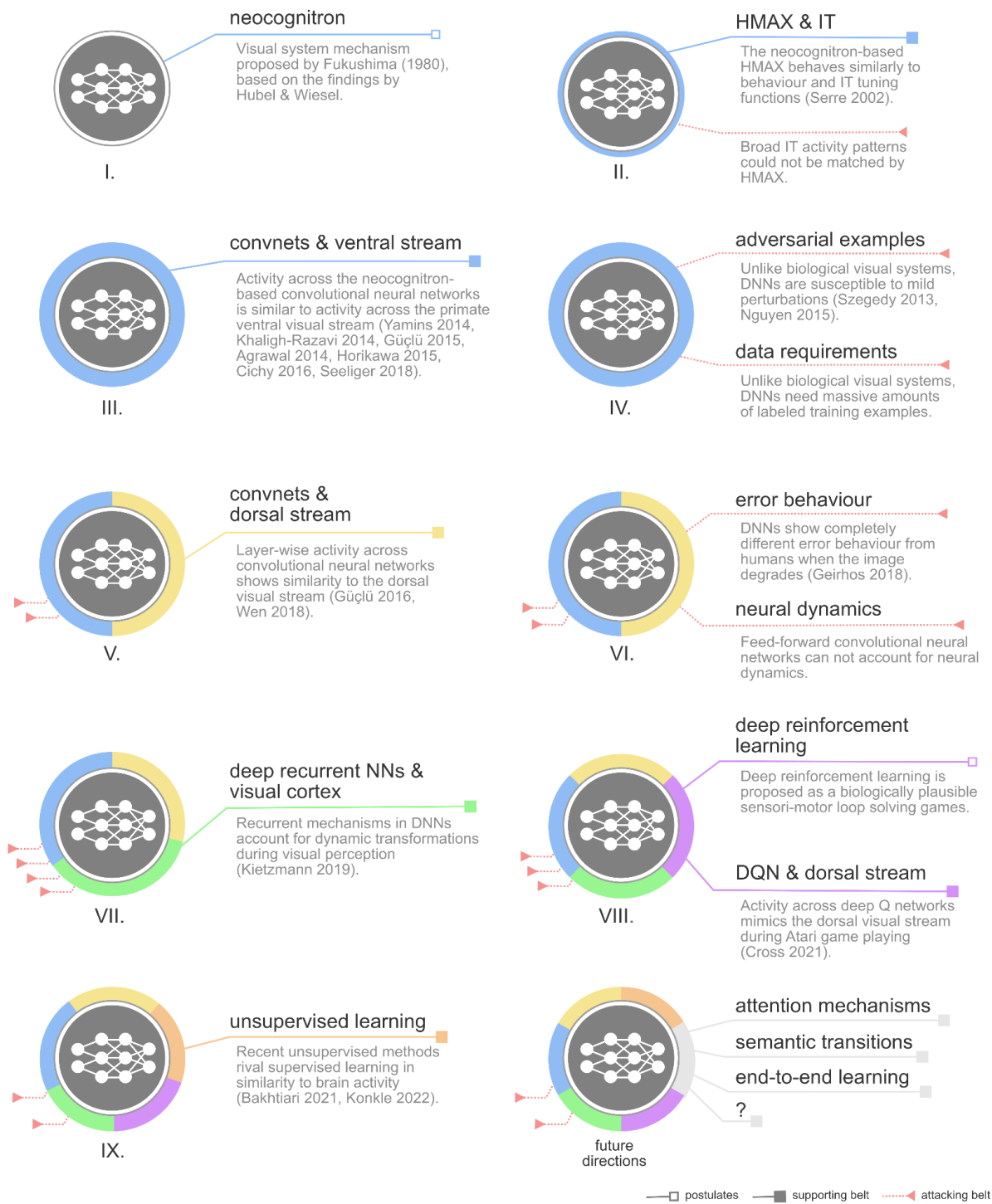

**Figure 3** | Summarized historical progression of the neuroconnectionism belt in the area of visual computational neuroscience. The figure shows how empirical and analytical findings are integrated into (or strengthen) the belt, and how objections are processed within a constructive, productive research programme. **I**. The neocognitron is suggested as a mechanism behind the seminal findings about simple and complex visual system cells by Hubel & Wiesel. It can learn and recognize increasingly abstract visual patterns through mechanisms that are similar to convolutions.  **II**. HMAX, a more powerful model similar to the neocognitron, is investigated closely for its similarity to neural activity. It matches IT tuning functions but does not match its broad activity patterns. It can not recognize objects in naturalistic photos. The neuroconnectionist belt

dwindles. **III**. Convolutional neural networks are successfully trained on large collections of naturalistic photos. A series of studies shows that their layer activity matches neural activity patterns along the primate ventral stream. This is the first time an image-computable and functional object recognition network matches activity patterns across the ventral visual system. These findings strengthen the neuroconnectionist belt and spawn a massive series of new studies and postulates. **IV**. With their susceptibility to adversarial attacks and the amounts of labeled training data they require, convolutional neural networks are shown to exhibit several behaviors that are not biologically plausible. **V**. Convolutional neural networks show similar layer activity to the dorsal stream, adding further experimental evidence strengthening the belt. **VI**. Adding to the growing list of biologically implausible mechanisms attacking the belt, it is shown that the error behavior during image degradation clearly diverges between humans and convolutional neural networks. Furthermore, feed-forward convolutional neural networks embody too simple mechanisms to cover neural dynamic observations beyond coarse rate coding. **VII**. The neural dynamics objection is resolved by the demonstration that dynamic transformations during visual processing can be captured if recurrence is added to DNNs. As recurrence is a biological property of the brain, this strengthens the neuroconnectionist belt. **VIII**. It is shown that activity across the dorsal stream during game playing matches activity in deep reinforcement learning networks, which implement a sensori-motor loop for the same game playing tasks. **IX**. Large numbers of labeled data are not required by unsupervised learning. Newer developments in this field are shown to match brain activity similarly well to supervised learning. As unsupervised learning is considered biologically plausible the resolution of this objection further strengthens the belt. **Future directions.** Attention mechanisms, semantic objectives and end-to-end learning in which networks are trained directly to match neural activity are all recent developments in ANNs. Experiments will tell which brain processes are better modeled by incorporating these elements.

*The evolution of the neuroconnectionism belt: a progressive generator of new insights.*

The belt of neuroconnectionism has considerably evolved in the past decade. By rapidly testing and expanding belt hypotheses, the community now commonly explores different architectures, datasets, objectives, and learning techniques across various experimental settings to test the resulting models for their alignment with brain and behavioral data.

One of the clearest examples of this progressive evolution in the field has been the change in how vision is modeled with ANNs in recent years (see Fig. 3 for an illustration of the historical progress of neuroconnectionism in vision). In the early 2010's, researchers focused a great deal of effort on comparing neural and behavioral data to what were, at that time, the state-of-the art in ANNs for vision, namely, deep convolutional neural networks trained on image classification tasks[15–18,20,21]. Over time, as researchers explored the successes and the failures of these models, the field has seen almost every component of these initial models updated, leading to new models that do a much better job of accounting for neurophysiological and psychological data. The earliest such update was a result of the recognition that the feed-forward nature of standard convolutional networks was both obviously incongruent with real neural anatomy, and functionally limiting. It was shown that recurrent networks were better at matching both behavioral data and neural activity patterns that occur with longer delays[19,33,76–80,82,150,174–177]. Additionally, researchers explored ways to improve the training datasets beyond computer vision benchmarks, including a more ecologically relevant selection of object categories[106], video data[178], embodiment[179,180] and goal-directed eye-movements[181]. Similarly, the use of supervised category training, which was always problematic from a neuroscience perspective - since humans do not need millions of labeled examples to learn - was shown to be unnecessary: recent work has shown that self-supervised techniques for training

ANNs lead to as good or better matches to neural representation and animal behavior[108,182,183]. Moreover, self-supervised training on video data can account for the distinction between the dorsal and ventral pathways in the brain[184] and self-supervised training on lower resolution inputs provides a better fit to the mouse visual cortex[185]. Other losses have also been explored, and researchers have found that dorsal stream vision can be explained both by control-based optimisation[35] and by self-motion related losses[186]. A similar evolution has occurred in models of the hippocampal formation and related networks, where initial architectures and losses have been replaced as the belt of the research program evolved. Early connectionist models largely focussed on attractor networks[187,188], which captured many interesting aspects of these circuits. But, with time, these models have evolved to incorporate additional architectural features and loss functions related to prediction and spatial integration, leading to better and better matches with a host of experimental results[189–191].

At the same time, as new ANN architectures have come out with more sophisticated attention mechanisms, such as transformers, more and more research has demonstrated that transformers trained in a self-supervised manner can effectively capture the representations observed in language areas of the brain[27,28], and other circuits, such as the mnemonic circuits of the medial temporal lobes. This leaves open the possibility for numerous other updates and explorations to existing models of vision and other senses, including the use of self-attention, and the use of multi-modal networks that combine linguistic inputs with vision or other sensory modalities[192,193].

In the above examples, neuroconnectionist models have provided state of the art models of behavioral and neural data, discovering new perspectives on the roles of known processes such as recurrence, attention and the dorsal stream. Importantly, fundamentally new theoretical insights and predictions about brain processing have also been derived, and some have been directly corroborated by in vivo experiments (see Box II).

---

**BOX II: PREDICTIONS & INSIGHTS GENERATED BY ANNs**

An important aspect of neuroconnectionism, as with all of computational modeling, is that it is not limited to providing accounts of existing data but can propose, test, and discover fundamentally new principles. In the following we describe some of such findings.

*ANNs have been used to derive novel predictions about neural selectivity.* Using optimization techniques that work at the level of the network input, it was shown that ANNs can be used to design novel stimuli that drive biological neurons in a highly specific manner, exceeding the firing rates observed using natural stimuli[164–166,194,195]. This causal manipulation provides an important extension to otherwise correlational tests of the alignment between models and the brain.

Another example of neuroconnectionism as a key for novel, and testable predictions about neural selectivity was provided by Bao et al.[40] who used ANNs to derive a theory of object selectivity in macaque IT. While certain parts of macaque IT had well-known selectivity profiles (e.g. face patches[6]), other parts remained uncharacterised "no-man's land", without any clear selectivity pattern. By using dimensionality reduction techniques on the late layers of an ANN trained on object classification, Bao et al. obtained a low-dimensional "object space" and found that responses of IT cells to a large set of objects aligned well with the axes of this space, including the

---

> previously uncharacterised no-man's land units. With this, the authors were able to provide a unified picture of IT organization as following a low-dimensional object space, as extracted from a deep network. See [31,44] for further such discoveries.
>
> *Neuroconnectionism has also yielded various new insights into the computational basis of neural information processing.* For example, based on the learning dynamics of ANNs, Saxe et al.[26] derived a mathematically tractable theory of semantic learning that recapitulates many empirical phenomena in human semantic development. Averbeck[196] proposed to explain adolescent changes in working memory by pruning in ANNs, and Rust & Jannuzi[197] used ANNs to derive insight into the memorability of images in humans. Nayebi et al.[185] studied the impact of combining embodiment and unsupervised learning on learning to explain the mouse visual cortex. What these examples have in common is that they use neuroconnectionist models to derive computational insights that may have been intractable without ANN simulations, due to the complexity of distributed neural codes.

These developments in the belt have raised new challenges in how we compare models and interpret results, requiring improvements in methodology. This has led to developments in quantifying the alignment of ANN and brain data[157,158,163,198,199], training networks end-to-end directly to match neural data[19,77,92–99], highlighting the importance individual variability across network instances[167], explicitly pitting networks against each other using "controversial stimuli"[200], integrating hierarchical[74] and temporal[19,150] aspects of information processing into model comparisons, performing detailed psychophysical experiments[140,144,150,201], using metamers to compare humans and models[202], and developing interpretable low-rank recurrent networks[203].

Altogether, these rapid developments of neuroconnectionism demonstrate that the underlying research programme is highly progressive. Researchers have repeatedly updated their models, altering the architectures, training objectives, and datasets to arrive at a progressively better account of neural information processing, including ANN-driven new insights and hypotheses. Although we are still far from final answers, and there are still important methodological issues[158,204–206], this illustrates how neuroconnectionist experiments, theory and methodology form a virtuous circle in which new empirical and theoretical results require improved methodology, which in turn allows for new results, allowing the field to keep progressing.

*Shortcomings as Signposts, not Roadblocks.*

At the same time, ANNs as models of brain function continue to be met with skepticism as differences with brain data and behavior remain. In the following, we consider prominent examples, asking whether the underlying controversies are *roadblocks* that preclude progress (and thereby may turn neuroconnectionism into a degenerating research programme), or whether they are better perceived as important *signposts*, which are pointing towards promising directions of improvement. In visiting these controversies, we highlight recent developments in the field that aim to solve them, suggesting that neuroconnectionism is a progressive research programme.

One of the main controversies surrounding ANNs in both cognitive science and artificial intelligence, concerns differences in the behavior of ANNs vs. biological brains in a variety of settings[53,56,207]. Although capable of impressive behavior, current ANNs do not model all aspects of behavior equally well: they generalize poorly[142], can be easily fooled[208–210] and behave differently from humans in many psychophysical settings[141,144,211]. These observations are important for the neuroconnectionist programme, as they point out instances where the current set of models exhibit shortcomings. Researchers are exploring a variety of avenues to address these shortcomings and improve the biological match of ANNs. As detailed above, attention mechanisms[212–214], recurrent mechanisms[19,76,80,82,139,174], foveation[215], and biology-inspired receptive fields in early layers[216–219] have been shown to improve different aspects of behavioral alignment with biology. Hence, the current shortcomings of ANNs in reproducing human behavior are not hindering the progress of neuroconnectionism. Rather, they are serving as inspiration for a multitude of new developments. They are real challenges that need to be taken seriously, but they are signposts, not roadblocks.

Another controversy focuses on computations on the level of symbolic manipulations in ANN vs biology. It has been argued that current ANNs, unlike humans, do not compute with symbols because they do not have appropriately structured representations, systematicity, dynamic variable binding and role-filler independence, which may ultimately cause tractability/scalability issues[129,220]. The field of neurosymbolic computations is a response to such criticisms[221–224]. Moreover, even the most ardent defenders of symbolism agree that ANNs should in principle be able to implement the symbolic systems we all desire[129,220,225]. Indeed, such a capacity is an immediate result of the universal approximation theorems that hold for neural networks[67,226]. Again, symbolic computations are not a roadblock for neuroconnectionism, they are signposts pointing to important directions of research that can be addressed from within the research program. The same goes for causal modeling[227] and related arguments.

Discussions surrounding ANNs as modeling framework are also based on their level of complexity. On the one hand, it is reasoned that (current) ANNs are *too abstract*, and that omitting many biological details renders them unable to adequately model neural processes. Indeed, current ANNs are vastly simpler than the breathtaking complexity of biological neural networks, which involves, to name a few examples, intricate dendritic computations[228,229], neurotransmitter dynamics, and communication via temporally fine-grained spikes[86,87,230]. At the same time, others deem ANNs *too complex* to provide a deeper understanding of cognition[54,231]. Indeed, the millions of parameters of ANNs constitute an important departure from classic models with only few interpretable parameters or experimental conditions. Critics ask, what is gained by modeling one complex system with another?

So are ANNs too simple or too complex? As mentioned in the introduction, although current ANNs are likely too simple, it is an empirical question at what level of detail cognitive systems are best understood. Instead of committing to a single level, the neuroconnectionist research programme is capable of modeling different levels of abstraction, which allows researchers to implement biological detail in a top-down fashion through the testing of hypotheses about which biological features are important for modeling cognitive phenomena[46,232]. We currently do not know which biological features are central for brain computations, and which are merely byproducts of evolution that could be equivalently implemented in other ways. Slowly testing the impact of more biological features in a hypothesis-driven fashion provides a way to simultaneously improve current ANNs and find out which biological features are important. At the same time, the comparably large number of ANN parameters is likely necessary to encode important domain knowledge, and to perform computations end-to-end from raw sensory input to behavioral output. Relatedly, the complexity of a model needs to be evaluated in relation to the complexity of the system being modeled. For example, the number of units in Alexnet (659k), a commonly used model in visual computational neuroscience, is equivalent to the number of neurons in a cube of cortex with an edge length of 2.3mm. Furthermore, the number of parameters is a poor proxy of model complexity in ANNs, and the degrees of freedom manipulated by the experimenter (i.e. architectural and learning hyperparameters) are orders of magnitude smaller.

In summary, current ANNs are both simple (e.g. in terms of architectural complexity), and complex (e.g. in terms of the number of adjustable parameters). But, to proponents of neuroconnectionism, this very balance is a feature and not a bug, since it allows for top-down testing of which biological details matter. As reviewed above, ANNs are actively moving towards increased realism in terms of architectures used, input datasets, as well as training objectives. The integration of biological detail, however, is driven as part of a hypothesis testing approach. This is importantly different from mirroring biology for its own sake, because biological details come at the cost of computational complexity and explanatory merit, giving good reason to try and discard the less relevant biological aspects.

Crucially, determining what level of complexity our models require in order to understand higher-level cognitive phenomena is itself an empirical project. Neuroconnectionism offers an ideal framework for studying which biological details are necessary and which are not, while at the same time being expressive enough to perform complex tasks. Put simply, *neuroconnectionist models live in a Goldilocks Zone between computational abstraction and biological detail*. They are sufficiently abstract to make them tractable and trainable (given the currently available methods and compute), but retain sufficient biological detail in their algorithmic structure to map them onto neural and behavioral data.

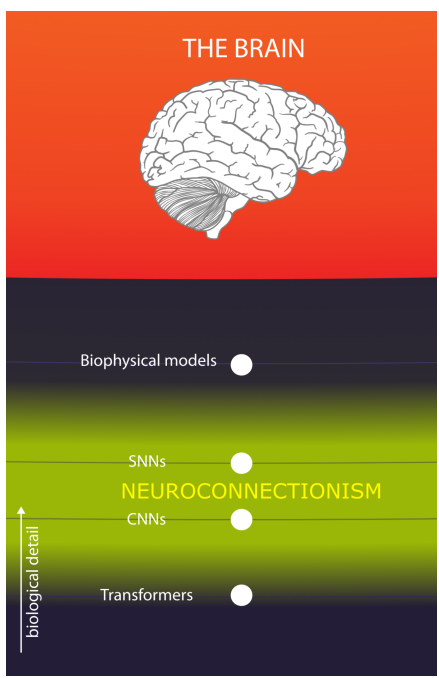

**Figure 4: The Goldilocks zone of biological abstraction.** Neuroconnectionism offers a coherent and computationally tractable framework to model cognition in which models vary in how much they abstract away from biology. This enables researchers to determine which biological details are needed to model essential aspects of brain computations. Hence, neuroconnectionism provides a goldilocks zone of biological abstraction, abstract enough to build large models capable of performing cognitive tasks, yet biologically detailed enough to test specific neuroscientific hypotheses.

*Conclusions*

In less than a decade, ANN modeling has gone from being fringe to a more central research tool in many parts of cognitive neuroscience, including vision, audition, and on to language and higher-level cognitive tasks. The underlying neuroconnectionism research programme has met both striking successes but also challenges. How should these contrasting results be interpreted? We have argued that the framework of progressive vs. degenerating Lakatosian research programmes is well suited to address this question. Of central importance is the view that neuroconnectionism is not a single theory or hypothesis. Instead, it is a research programme composed of many different auxiliary hypotheses and research directions in the belt, each sharing the same core approach to model cognitive phenomena through distributed neural communications in ANNs. Based on this view, we showed that current challenges represent important signposts that aid further progress, rather than roadblocks. Indeed, many such challenges have sparked vibrant new research directions. Together with the growing body of studies demonstrating good agreement between connectionist systems and brain data, these observations illustrate that the neuroconnectionism research programme is highly progressive. Importantly, traditional approaches based on simple human-interpretable concepts face significant challenges that can be overcome by the addition of neuroconnectionist models to the researcher's toolbox. These include explanations of neural data collected under more natural and varied stimulus paradigms, explanations of lower-level sensory processes and their transition to higher-level conceptual information, explanations of behavioral effects based on raw sensory input, and learning. Another argument supporting the use of ANNs in neuroscience is that theoretical computer science has taught us that complex systems often cannot be simplified to predict their behavior: the only way to know the outcome of processing is to actually run the system (or a good enough ANN model of it).

One reason for the success of ANNs may be that they provide a useful level of granularity for cognitive neuroscience. They provide a sufficiently abstract view on neural processes while being able to incrementally test which biological detail is needed in a top-down, hypothesis-driven manner. The hyperparameters of the training procedure thereby form the effective degrees of freedom of the experimenter, not the number of parameters in the respective network. On the other hand, ANNs are high-dimensional enough to encode domain knowledge, enabling them to be grounded in sensory input while performing comparatively well on a set of behavioral tasks.

While we have here focused on the progressive nature of the research programme, we do not mean to imply that the field is anywhere near to successfully explaining cognition. Neuroconnectionism is still in its infancy, and hence knowing where it fails is equally important as knowing where it works. In addition to addressing current challenges, future work will need to incorporate many currently missing aspects of lower and higher cognition. Focusing purely on the case of visual processing for the sake of argument, these include (i) multi-task networks that can explain our ability to perform multiple, sometimes highly abstract tasks based on sensory evidence[233], (ii) a focus on embodiment rather than treating networks as artificial brains in vats[179,185,186,234], (iii) data efficient and continuous learning using biologically realistic learning rules and inductive biases, (iv) an answer to how symbolic reasoning is implemented in neuroconnectionist models, (v) better methods for unsupervised, multimodal learning, (vi) better modeling of cognitive development, (vii) the integration of multiple memory systems, and (viii) learning in social context (learning from each other). The fact that all these aspects are currently missing but can in principle be implemented in ANNs perfectly illustrates both the current limitations and strong potential of the neuroconnectionist approach. Rather than feeling threatened by the possibility of a new connectionist winter, the neuroconnectionist community should therefore continue to welcome criticism and limitations as they point the way towards new insights. Critics of neuroconnectionism on the other hand should not regard every shortcoming of the current set of networks as a failure of the entire research programme. Time will tell whether neuroconnectionism can deliver on its promises to explain the emergence of cognitive phenomena, behavior and neural data from bio-inspired, yet simple distributed coding principles. For now, it remains a highly progressive and therefore exciting research programme that welcomes critical signposts to guide the way.

*Acknowledgements*

The authors acknowledge support by SNF grant n.203018 (Doerig), the ERC stg grant 101039524 TIME (Kietzmann), and the Max Planck Research Group grant of Martin N. Hebart (Seeliger).